\documentstyle[12pt]{article}                                         

\begin{document}                                                      

\baselineskip 22pt                                                    
\hfill hep-th/9603197 \\
       
\begin{center}                                                        
          
{\Large \bf Geometrical Interpretation of BRST Symmetry in 
Topological Yang-Mills-Higgs Theory}

\vspace{1.0cm}                                                        
          
{\large Chang-Yeong {\sc Lee}}$ \: {}^{\dag}$  \\       
{\it Department of Physics, Sejong University, Seoul 143-747, KOREA}\\

\vspace{1.0cm}

{\large \bf Abstract} \\                                                  
\end{center}                                                          
We study topological Yang-Mills-Higgs theories in two and three dimensions 
and topological Yang-Mills theory in four dimensions in a unified framework
of superconnections. In this framework, we first show that a classical 
action of topological Yang-Mills type can 
provide all three classical actions of these theories 
via appropriate projections.  
Then we obtain the BRST and anti-BRST transformation rules encompassing
these three topological theories from an extended definition of curvature
and a geometrical requirement of Bianchi identity. This is an extension of
Perry and Teo's work in the topological Yang-Mills case. 
Finally, comparing this result with our previous treatment 
in which we used the ``modified horizontality condition",
we provide a meaning of Bianchi identity from the BRST symmetry viewpoint
and thus interpret the BRST symmetry in a geometrical setting. 
\\
 
\vfill

\noindent
PACS number(s): 02.40.-k,  11.15.-q, 11.30.-j\\
\hbox to 12cm{\hrulefill}  \\
${}^{\dag} \; $ Electronic-mail: leecy@hep.sejong.ac.kr\\
\thispagestyle{empty}
\pagebreak

\noindent
{\large \bf I. Introduction}\\

Soon after Witten \cite{wtt} constructed topological Yang-Mills theory 
 to generate the Donaldson
invariants of smooth four-manifolds, 
Baulieu and Singer \cite{bs} showed that Witten's 
topological quantum action could be obtained 
by gauge fixing the classical topological action
\begin{equation}
I_{4} = \int_{M_{4}} {\rm Tr} \ F \wedge F
\label{c1}
\end{equation}
in the BRST quantization scheme.
Then Perry and Teo \cite{pt} argued that the asymmetry of BRST 
transformation rules appeared in  
the Baulieu and Singer's work 
was caused by treating only
the BRST symmetry \cite{brst}, and not the anti-BRST symmetry \cite{cf}.
And they obtained symmetric BRST and anti-BRST transformation rules
by treating them on an equal footing.
They further identified the difference between the ordinary Yang-Mills theory
and the topological Yang-Mills theory as follows.
In ordinary Yang-Mills theory, one can impose  
the so-called ``horizontality condition" 
to find BRST symmetry, and this is tantamount to requiring the
vanishing Yang-Mills field strength (curvature) 
along the unphysical directions of ghosts.
In the topological case, one can not impose this condition of vanishing
field strength along the unphysical
directions, 
and one can only impose the Bianchi identity in the
ghosts-included extended space.
 
Parallel 
to this development, 
topological Yang-Mills-Higgs actions in two and three dimensions were
also constructed.  
Following the Baulieu and Singer's approach,
Baulieu and Grossman \cite{bg} found the topological action for 
magnetic monopoles  by gauge fixing the following classical  
action in three dimensions  
\begin{equation}
I_{3} = \int_{M_{3}} {\rm Tr} \ F \wedge D \phi .
\label{c2}
\end{equation}
Two dimensional case was studied by Chapline and Grossman \cite{cg}
by gauge fixing the following two dimensional classical action
\begin{equation}
I_{2} = \int_{M_{2}} {\rm Tr} \left( F [ \Phi^{\dag} , \Phi] -  
   D \Phi^{\dag} \wedge D \Phi \right),
\label{c3}
\end{equation}
and the quantized theory turned out to be connected
to the theory of vortices and knots.

In our previous work \cite{hl2}, we investigated 
the BRST/anti-BRST symmetry of the above topological 
Yang-Mills-Higgs theory in 
two and three dimensions by modifying the horizontality 
condition such that it could take care of the topological symmetry
in addition to the ordinary Yang-Mills gauge symmetry.
This work was done in the superconnection framework 
so that the scalar and vector gauge fields 
were treated on the same footing
as a connection. Thereby we could find the BRST/anti-BRST
transformation rules of the scalar and vector gauge fields at once without
doing separate calculations.

In this paper, we investigate the BRST/anti-BRST symmetry of these 
theories through the ``Bianchi identity" 
in the same superconnection framework. This was motivated by a question 
that is whether the Perry and Teo's work could be extended to the
topological Yang-Mills-Higgs case in which additional ghosts for the
scalar field appears. 
Thus the superconnection framework became a very natural testing 
ground for this idea. The result is the affirmative. 
By comparing these two approaches,
we can further provide a meaning of the Bianchi identity 
in the extended space
 from the BRST symmetry view point. This in turn allows a
geometrical interpretation of the BRST/anti-BRST symmetry  
in the extended space.
 
In section II, we show that the classical actions, 
$I_{4}, \; I_{3}, \; I_{2},$ can be obtained from a classical action 
of topological Yang-Mills type written in superconnection language
by appropriate projections depending on the dimensions of spaces 
to which corresponding theories belong.
In section III, we find the BRST/anti-BRST transformation rules of
the topological Yang-Mills-Higgs theory from the Bianchi identity 
in the extended space and an extended definition of  
curvature in the superconnection formalism.
In section IV, we compare our present work with the ``horizontality 
condition" approach which we adopted in our previous work. 
   From the comparison of these two approaches, we
provide a geometrical meaning to the BRST symmetry in the topological
Yang-Mills-Higgs theory.
Section V consistutes the conclusion.
\\

\noindent
{\large \bf II. Classical topological action with superconnections} \\

In 1982, Thierry-Mieg and Ne'eman \cite{tmn} constructed a 
generalized system of connections with arbitrary form degrees 
hinted from the old idea of Cartan's 
integrable system \cite{crt}, while they studied a generalized gauge
theory possessing an internal supersymmetry. 
In mathematics, a smiliar concept was introduced by 
Quillen in 1985 \cite{qui} under a notion of superconnections, independently
to Thierry-Mieg and Ne'eman's work.
Then Ne'eman and Sternberg \cite{ns} used the Quillen's 
superconnection concept
to study the Higgs mechanism where the Higgs field occurs as the zero-th
order part of the superconnection. This work is much easier for physicists 
to understand the superconnection concept and also shows that the 
superconnection is not much different from the generalized connection of
Thierry-Mieg and Ne'eman except for the existence of zero-th order connection.
In this paper, we thus follow the Ne'eman and Sternberg 
presentation of superconnections.
In general, the superconnection has all orders with 
odd degree forms as its even part and even degree forms as its odd part. 
However, in this paper we shall deal with superconnections which 
contain zero and one forms only since the theories we are dealing with
have the scalar and vector gauge fields only.
Now, we write down our superconnection as 
\begin{equation}
{\cal J}  = \left( \begin{array}{cc} A & i \Phi \\
  i \Phi^{\dag}  & A  \end{array} \right)
\label{c4}
\end{equation}
where $A, \Phi$ are Lie algebra valued one form and zero form, respectively.
The multiplication rule among the elements of total $Z_{2}$-graded  
``superspace" is given by \cite{ns}
\begin{equation}
\left( \begin{array}{cc}A & C\\ D & B \end{array} \right) 
\left( \begin{array}{cc}A' & C' \\ D' & B' \end{array} \right) =
\left( \begin{array}{cc} A \wedge A' + (-1)^{ \mid {\rm D'} \mid } 
C \wedge  D' & A \wedge C' + 
(-1)^{ \mid {\rm B'} \mid } C \wedge B' \\
 (-1)^{ \mid {\rm A'} \mid } D \wedge A' + 
 B \wedge D' & (-1)^{ \mid {\rm C'} \mid }
 D \wedge C' + B \wedge B' \end{array} \right) 
\label{c5}
\end{equation}
where $A,\cdots, A', \cdots $ are matrices of differential forms, and
$\vert A' \vert, \vert B' \vert, \vert C' \vert, \vert D' \vert $ denote 
form degrees of $A', B', C', D',$  respectively. 

The ``super" curvature is defined from superconnection as 
\begin{equation}
{\cal F} = {\bf d} {\cal J} + {\cal J} {\cal J} ,
\label{c6}
\end{equation}
where {\bf d} denotes a one form differential operator given by 
$ {\bf d} = \left( \begin{array}{cc} d & 0 \\ 
                            0 & d \end{array} \right) $  
with $ d$ denoting the ordinary one form exterior derivative 
times a unit matrix. From here on, we shall use the term 
curvature instead of ``super" curvature for brevity.
Written in the component form, the curvature is given  by
\begin{equation}
{\cal F}  = \left( \begin{array}{cc} F - \Phi
\Phi^{\dag}   & i D \Phi \\
 i D \Phi^{\dag} &    F - \Phi^{\dag} \Phi \end{array} \right)
\label{c7}
\end{equation} \\
where $F=dA + A \wedge A $ and $D \Phi =d\Phi + A \Phi - \Phi A $.
Now, we claim our classical topological action as
\begin{equation}
I = \int_{M} {\rm GTr} \; {\cal FF}
\label{c8}
\end{equation}
and we explain what ``GTr" means below.
In general, we can write down ${\cal F}$ as
${\cal F}  = \left( \begin{array}{cc} {\cal F}_{ev} 
    & ({\cal F}_{od})_1 \\
 ({\cal F}_{od})_2 &  {\cal F}_{ev} \end{array} \right) $, thus
${\cal FF} $ can be written as
\begin{equation}
{\cal FF} = \left( \begin{array}{cc} ({\cal F}_{ev})^2 - 
   ({\cal F}_{od})_1  ({\cal F}_{od})_2
    & {\cal F}_{ev} ({\cal F}_{od})_1 + ({\cal F}_{od})_1 {\cal F}_{ev}
    \\
 ({\cal F}_{od})_2 {\cal F}_{ev} +{\cal F}_{ev} ({\cal F}_{od})_2
 &  ({\cal F}_{ev})^2 - ({\cal F}_{od})_2 ({\cal F}_{od})_1 
 \end{array} \right)  .
\label{c9}
\end{equation}
In four dimensions, only $  ({\cal F}_{ev})^2 $ term can contribute since
${\cal F}_{ev}$ is either two form or zero form. Thus we take the ordinary
trace for ``GTr" in order to get a meaningful result. In three dimensions,
only $  {\cal F}_{od} {\cal F}_{ev} $ type terms can contribute since 
$ {\cal F}_{od}$ terms are one forms. And in this case we take ``GTr" as
taking the ordinary trace after the addition of the odd parts, and 
we denote this as ``QTr" following the 
notation of the queer trace defined in Ref. \cite{bz}.
In two dimensions, two types of terms can contribute,
$ ({\cal F}_{ev})^2$ and ${\cal F}_{od} {\cal F}_{od}$. However, only 
$ {\cal F}_{od} {\cal F}_{od}$ type terms have second derivative terms
and we take ``GTr" such that these terms do not vanish. 
Thus we take supertrace in the two dimensional case.
Given this rule, the classical topological action (\ref{c8}) becomes \\
(a) in four dimensions
\begin{eqnarray}
I & = & \int_{M_{4}} {\rm Tr} \; {\cal FF} \nonumber \\
 & = & 2 \int_{M_{4}} {\rm Tr} \ F \wedge F ,
\label{c10}
\end{eqnarray}
(b) in three dimensions
\begin{eqnarray}
I & = & \int_{M_{3}} {\rm QTr} \; {\cal FF} \nonumber \\
 & = & 4i \int_{M_{3}} {\rm Tr} \ F \wedge D \phi \label{c11}
\end{eqnarray}
where $ \phi = \frac{1}{2} ( \Phi^{\dag} + \Phi ), $  \\
(c) in two dimensions
\begin{eqnarray}
I & =& \int_{M_{2}} {\rm STr} \; {\cal FF} \nonumber \\
  & = & 2 \int_{M_{2}} {\rm Tr} \ ( F [\Phi^{\dag}, \Phi] -
           D \Phi^{\dag} \wedge D \Phi ) \label{c12}
\end{eqnarray}
where we used the anticommuting property of one form $D \Phi$.
In this way, we retrieve all three classical actions of 
topological Yang-Mills-Higgs theory in Refs. \cite{bs,bg,cg}.
\\

\noindent
{\large \bf III. Curvature, Bianchi identity, and BRST/anti-BRST symmetry}\\

In the geometrical BRST quantization scheme, 
the base space is extended in such a way that   
the ghost/antighost sector can be constructed on the extended space. 
This is done by adding a doubled fiber bundle structure 
to the base manifold to represent unphysical (ghost/antighost) directions.
This scheme was first developed by Thierry-Mieg and Ne'eman \cite{tm,ntm} 
with principal fiber bundle structure yielding 
the BRST symmetry (the ghost direction) only.
Then it was further developed to yield  the BRST and anti-BRST symmetry 
together by including the antighost direction also -- 
a doubled fiber bundle structure \cite{qrs}.
In this scheme, the ghost/antighost fields are obtained from
the gauge field by replacing its spacetime leg $dx^{\mu}$ 
with $dy^N$ ($d \bar{y}^N$) where $y, \  \bar{y}$ represent the 
fiber coordinates in a doubled fiber bundle \cite{tmn,tmn2,btm,ln}.
 If one does not like the interpretation of this extended
fiber bundle approach, one can
take the superspace interpretation given in Refs. \cite{botn,bl}, 
whose view was
taken in Perry and Teo's work \cite{pt}. In the superspace approach, the
fiber coordinates  $y, \ \bar{y}$ are replaced by
a set of anticommuting variables $\theta$ and $\bar{\theta}$
which represent the coordinates of the abstract superspace extended from
the spacetime basemanifold.
However, the resultant
BRST/anti-BRST transformation
rules are exactly the same whichever approach one uses, and thus we will
be careless of the subtleties among the two approaches.
In the ordinary Yang-Mills theory, the BRST/anti-BRST symmetry is 
obtained from a condition which sets that the ordinary curvature 
equals to the extended curvature, which is tantamount to 
setting the curvature
components containing vertical (fiber) directions zero, thus 
only the horizontal components
of curvature (physical Yang-Mills field strength) 
in the extended space survive.
For this reason, people gave the name ``horizontality condition" \cite{btm}
to this condition.
In this paper, we denote objects in the extended space with tildes.

Following this geometrical BRST scheme, we first extend the 
superconnection as 
\begin{equation}
\widetilde{\cal J} ={\cal J} + {\cal C} + \overline{\cal C}
\label{c13}
\end{equation}
where ${\cal C}$ and $\overline{\cal C}$ are the first generation ghost and
antighost for ${\cal J}$, which  
are given by 
\begin{equation}
{\cal C} = \left( \begin{array}{cc} c   & 0 \\
 0 &   c \end{array} \right), \; \; \; \; \overline{\cal C} =
\left( \begin{array}{cc} \bar{c}   & 0 \\
 0 &  \bar{c} \end{array} \right).
\label{c14}
\end{equation}
Here $ \ c, \ \bar{c}$ denote 
$ \ c= A_N d y^N, \ \bar{c}=\bar{A}_N d \bar{y}^N$,
and represent the ghost and antighost fields, respectively.
In this extended space, the curvature is given by 
\begin{equation}
\widetilde{\cal F} = \widetilde{\bf d} \ \widetilde{\cal J} +
  \widetilde{\cal J}  \widetilde{\cal J}
\label{c15}
\end{equation}
where 
\begin{equation}
\widetilde{\bf d} = {\bf d} + {\bf s} + \bar{\bf s}.     
\label{c16}
\end{equation}
Here, ${\bf s}$ and $\bar{\bf s}$ denote one form exterior 
derivative operators
acting on ghost and antighost directions 
expressed in superconnection language, 
as ${\bf d}$ does in spacetime directions.
Now, following the spirit of Refs. \cite{bs,pt}, we identify the
curvature components in unphysical directions with new fields. 
One main difference is that here we have the first generation ghost and
antighost fields which are one forms in the extended space 
(having only $dy$ or $d \bar{y}$), because 
in the superconnection formalism we have one form curvature components
due to the scalar field.
\begin{equation}
\widetilde{\cal F} = \left( \begin{array}{cc} 
F - \Phi \Phi^{\dag} + \psi + \bar{\psi} + m + \lambda + \bar{m}
& i ( D \Phi + \xi + \bar{\xi}) \\ i ( D \Phi^{\dag} + \xi^{\dag} + 
\bar{\xi}^{\dag}) & 
F - \Phi^{\dag} \Phi + \psi + \bar{\psi} + m + \lambda + \bar{m}
\end{array} \right)
\label{c17}
\end{equation}
where $\psi, \  \bar{\psi}, \  m, \ \lambda, \ \bar{m}$ are
the first and second
generation ghost and antighost fields for the two form curvature $F$,
and $\xi, \ \bar{\xi}$ are the 
first generation ghost and antighost fields for
the one form curvature $D \Phi$:
\begin{eqnarray}
\psi & = & {\cal F}^{1}_{\mu N} dx^{\mu} dy^N \nonumber \\
 \bar{\psi} & = & {\cal F}^{-1}_{\mu N} dx^{\mu} d \bar{y}^N \nonumber \\
  m      & = & {\cal F}^{2}_{M N} dy^M dy^N   \nonumber \\
\lambda & = & {\cal F}^{0}_{M N} dy^M d \bar{y}^N \label{c18} \\
\bar{m} & =& {\cal F}^{-2}_{M N} d \bar{y}^M d \bar{y}^N \nonumber \\
\xi     & =& {\cal F}^{1}_{N} dy^N \nonumber \\
\bar{\xi} & = & {\cal F}^{-1}_{N} d \bar{y}^N \nonumber
\end{eqnarray}
where upper indices $1,-1,2, {\it etc.},$ 
represent ghost numbers.
For instance,
$\psi$ has ghost number 1 and $ \bar{\psi}$ has ghost number $-1$.
   
The curvature in the extended space should also satisfy the Bianchi identity,
\begin{equation}
\widetilde{\bf d} \widetilde{\cal F} +
 [ \widetilde{\cal J}, \widetilde{\cal F} ] = 0 .
\label{c19}
\end{equation}
Thus we have two conditions:  
(a) we have to equate Eq.(\ref{c15}) with Eq.(\ref{c17}), and
(b) the Bianchi identity Eq.(\ref{c19}). 
The rules of BRST/anti-BRST
symmetry are obtained from these two conditions. 
The BRST/anti-BRST transformation rules for the components 
of the extended superconnection
$\widetilde{\cal J}$ are given by the first condition.
\begin{eqnarray}
{\rm even \ \ part:} & & sA +dc +cA +Ac = \psi, \nonumber \\
 & & \bar{s} A + d \bar{c} + \bar{c} A + A \bar{c} = \bar{\psi}, \nonumber \\
 & & sc + cc = m,  \nonumber \\
 & & \bar{s} \bar{c} + \bar{c} \bar{c} = \bar{m},  \label{c20} \\
 & & s \bar{c} + \bar{s} c + c \bar{c} + \bar{c} c = \lambda, \nonumber \\
{\rm odd \ \ part:} & & s \Phi + c \Phi - \Phi c = \xi, \nonumber \\
   & & \bar{s} \Phi + \bar{c} \Phi - \Phi \bar{c} = \bar{\xi}. \nonumber
\end{eqnarray}
The second condition, the Bianchi identity, gives the BRST/anti-BRST
 transformation rules 
for the compononents of the extended curvature $\widetilde{\cal F}$.
\begin{eqnarray}
{\rm even \ \ part:} & & s\psi + dm +Am +c\psi -mA -\psi c = 0, 
       \nonumber \\
   & & \bar{s} \bar{\psi} + d \bar{m} + A \bar{m} + \bar{c} \bar{\psi}
        - \bar{m} A - \bar{\psi} \bar{c} = 0, \nonumber \\
   & & s \bar{\psi} + \bar{s} \psi + d \lambda + A \lambda + c \bar{\psi}
      +  \bar{c} \psi - \lambda A - \psi \bar{c} - \bar{\psi} c = 0 ,
          \nonumber \\
    & & sm +cm -mc = 0, \nonumber \\
  & & \bar{s} \bar{m} + \bar{c} \bar{m} - \bar{m} \bar{c} = 0, \nonumber \\
  & & s \lambda + \bar{s} m + c \lambda + \bar{c} m -m \bar{c} 
         - \lambda c =0, \label{c21} \\
  & & s \bar{m} + \bar{s} \lambda + c \bar{m} + \bar{c} \lambda 
        - \lambda \bar{c} - \bar{m} c = 0, \nonumber \\
{\rm odd \ \ part:} & & s \xi + c \xi + \Phi m - m \Phi + \xi c = 0, 
          \nonumber \\
          & & \bar{s} \bar{\xi} + \bar{c} \bar{\xi} + \Phi \bar{m}
              - \bar{m} \Phi + \bar{\xi} \bar{c} = 0, \nonumber \\
    & & s \bar{\xi} + \bar{s} \xi + c \bar{\xi} + \bar{c} \xi + \Phi \lambda
        - \lambda \Phi + \xi \bar{c} + \bar{\xi} c = 0. \nonumber
\end{eqnarray}
As usual, we have to introduce auxiliary fields to completely fix the 
BRST/anti-BRST transformation rules.
We first define auxiliary fields   
\begin{eqnarray}
s \bar{c} & = & b , \nonumber \\
s \bar{\psi} & = & - \kappa , \nonumber \\
s \lambda & = & \eta , \label{c22} \\ 
s \bar{m} & = & \bar{\eta}, \nonumber \\
s \bar{\xi} & = & \zeta, \nonumber
\end{eqnarray}
then we get from Eqs.(\ref{c20},\ref{c21})
\begin{eqnarray}
\bar{s} c & = & -b - [c, \bar{c} ] + \lambda , \nonumber \\
\bar{s} \psi & = & \kappa - D \lambda - [c, \bar{\psi} ] - [\bar{c}, \psi ],
          \nonumber \\ 
\bar{s} m & = & - \eta - [ c, \lambda ] - [\bar{c}, m], \label{c23} \\
\bar{s} \lambda & = & - \bar{\eta} -[c, \bar{m}] -[\bar{c}, \lambda], 
      \nonumber \\
\bar{s} \xi & = & - \zeta - [ \Phi, \lambda ] -[c, \bar{\xi}] -[\bar{c},\xi ].
\nonumber
\end{eqnarray}
The nilpotency of BRST/anti-BRST transformation operators, $s^2=\bar{s}^2=0$,
determines all the rest  
\begin{eqnarray}
sb=0, & \; \; \; & \bar{s} b =[b, \bar{c}] - \bar{\eta}, \nonumber \\
s \kappa =0, &  & \bar{s} \kappa = - [ b, \bar{\psi}] + D \bar{\eta} 
     -[\bar{c}, \kappa ] + [\bar{m}, sA], \nonumber \\
s \eta = 0, &  & \bar{s} \eta = [b, \lambda] -[c, \bar{\eta}] 
        -[\bar{c}, \eta] - [\bar{m}, sc ], \label{c24} \\  
s \bar{\eta} = 0, &  & \bar{s} \bar{\eta} = [b, \bar{m} ] -[\bar{c},
        \bar{\eta} ] , \nonumber \\
s \zeta = 0,  &  & \bar{s} \zeta = [b, \bar{\xi} ] - [\Phi, \bar{\eta} ]
      -[ \bar{c}, \zeta ] - [ \bar{m}, s \Phi ] , \nonumber
\end{eqnarray}
where $ sA, \ sc, \ s \Phi$ are given in Eq.(\ref{c20}).
In Eqs.(\ref{c23},\ref{c24}), $[ \ , \ ] $
denotes a graded commutator. For instance, 
$ [ c, \bar{c} ]= c \bar{c} +  \bar{c} c$, and $[b, \bar{c}]= b \bar{c} -
  \bar{c} b \ $ since $ c, \bar{c}$ are anticommuting fields and 
$b$ is a commuting field. In this way, we obtain all the 
transformation rules of the BRST/anti-BRST symmetry
in topological Yang-Mills-Higgs theory 
obtained in Refs. \cite{pt,bg,cg}.      
\\

\noindent
{\large \bf IV. Comparison with the ``horizontality condition" approach}\\

The BRST symmetry of the ordinary Yang-Mills theory can be obtained 
from the so-called horizontality condition \cite{btm}.  
On the other hand, the BRST symmetry of topological Yang-Mills theory 
is not obtained through 
a strict application of the horizontality condition, rather it was 
obtained thorough a modified definition of curvature and the Bianchi
identity in the extended space \cite{pt}. 
 In Ref. \cite{hl2}, we modified the horizontality
condition such that it could yield the complete BRST symmetry of topological
Yang-Mills-Higgs theory.
The rationale of this modified horizontality condition is the following.
In the ordinary Yang-Mills case, the curvature in the extended space has 
vanishing components along the vertical directions which represent gauge fiber
orbits of classical gauge symmetry, and this fact is expressed as the
horizontality condition
\begin{equation}
\widetilde{F}=\widetilde{d} \widetilde{A} + \widetilde{A}\widetilde{A}
     = dA + A A = F
\label{c25}
\end{equation}
where $ \widetilde{d} = d + s + \bar{s} $ and 
$ \widetilde{A} = A + c + \bar{c}$.  
In the topological case, we have larger symmetry than the gauge symmetry
and this extra symmetry also has to be gauge fixed. 
That means we need extra ghosts besides the 
ordinary ones ($c, \ \bar{c}$) orginated from the 
gauge symmetry. 
Hence we modify the horizontality condition by adding ``permissible"
ghosts to the extended curvature 
\begin{equation}
\widetilde{F}_{T} = F,  \hspace*{1cm} {\rm where} \hspace*{1cm} 
\widetilde{F}_{T}=\widetilde{F} + \widetilde{F}' ,
\label{c26}
\end{equation}
such that $  \widetilde{F}' $ consists of ghosts (antighosts) only and
satisfies the nilpotency of BRST symmetry, $ s^2 \widetilde{F}' ( =
\bar{s}^2 \widetilde{F}') =0 $. 
Also this $\widetilde{F}'$ has to be choosen in such a way that it respects
$ s^2 \widetilde{A} ( = \bar{s}^2 \widetilde{A} ) = 0$.
Through this way we can obtain the correct BRST/anti-BRST symmetry 
of topological Yang-Mills theory of Ref. \cite{pt}. 
What we explained so far is for the topological Yang-Mills case, not
including the Higgs field. 
In order to encompass the topological Yang-Mills-Higgs case \cite{bg,cg},
we carried out the same procedure in the superconnection framework
in our previous work \cite{hl2}: 
\begin{equation}
\widetilde{\cal F}_{T} = {\cal F},  \hspace*{1cm} {\rm where} \hspace*{1cm}
\widetilde{\cal F}_{T}=\widetilde{\cal F} + \widetilde{\cal F}' 
\label{c26a}
\end{equation}
with $ {\cal F}, \ \widetilde{\cal F}$ given by Eqs.(\ref{c7},\ref{c15}),
respectively, and \[  \widetilde{\cal F}' = - \left( \begin{array}{cc}
 \psi + \bar{\psi} + m + \lambda + \bar{m}
& i ( \xi + \bar{\xi}) \\ i (  \xi^{\dag} + \bar{\xi}^{\dag}) &
 \psi + \bar{\psi} + m + \lambda + \bar{m} \end{array} \right). \] 

Now, comparing the above approach with the Bianchi identity approach
that we carried out in this paper, we note two things. 
First, the newly defined components of the extended curvature in the Bianchi
identity approach correspond to the additional curvature $\widetilde{F}'$
(or $\widetilde{\cal F}'$) 
in the modified horizontality condition approach with a negative sign.  
Second, the requirement of the Bianchi identity for the 
newly defined curvature in Eq.(\ref{c17})   is
replaced with the BRST/anti-BRST nilpotency condition on the extra 
curvature $\widetilde{F}'$ (or $\widetilde{\cal F}'$)  in the 
modified horizontality condition approach. 
Now the first observation tells us that the newly defined curvature 
components ($ \psi, \ \bar{\psi}, \  m, \  \lambda, \  
\bar{m}, \ \xi, \ \bar{\xi}$) in the 
extended space given in Eq.(\ref{c17}) represent the existence of topological
symmetry other than the ordinary gauge symmetry which is taken care of
by the ghost sector of the extended connection $ \widetilde{A} $ 
(or $ \widetilde{\cal J}$). 
The second observation tells us that the 
Bianchi identity in the extended space is simply another 
expression of the BRST/anti-BRST nilpotency 
condition for the extra ghost/antighost fields 
which appear as the new curvature components.
In fact, in the Bianchi identity approach, the Bianchi identity 
in the extended space also implies this point: 
\begin{equation}
\widetilde{d} \widetilde{F} + [ \widetilde{A}, \widetilde{F} ]
= 0, \hspace*{.5cm} {\rm where} \hspace*{.5cm}
 \widetilde{F} = \widetilde{d} \widetilde{A} + \widetilde{A} \widetilde{A}
\label{c27}
\end{equation}
The above Bianchi identity is valid due to the nilpotency of 
the extended exterior derivative
$\widetilde{d}= d + s + \bar{s} $,
 and the nilpotency of $\widetilde{d}$ implies the nilpotency of
the BRST symmetry, $ s^2 = \bar{s}^2 = 0$.
Also in the Bianchi identity approach, the extended curvature
is defined to contain
the extra curvature  $\widetilde{F}'$ (or $ \widetilde{\cal F}'$) of
the modified horizontality condition approach. 
Thus in the Bianchi identity approach one does not require the 
horizontality condition
and instead the new degrees of freedom in the unphysical 
directions are allowed.
For instance for the topological Yang-Mills case, we define the
extended curvature $ \widetilde{F}$ on which we do not impose
the horizontality condition as follows:
\begin{equation}
 \widetilde{F}=  \widetilde{d} \widetilde{A} + \widetilde{A} \widetilde{A}
  = F + \psi + \bar{\psi} + m + \lambda + \bar{m} .
\label{c28}  
\end{equation} 
Since $\widetilde{A}= A + c + \bar{c}$ and $\widetilde{d}= d + s + \bar{s}$,
the above definition of $ \widetilde{F}$ allows the gauge field $A$ 
to carry extra symmetry represented by the ghost $ \psi$
as we have seen in Eq.(\ref{c20}).
And the symmetry property of the additional ghosts 
($ \psi, \  \bar{\psi},$ {\it etc.}) 
is constrained by the Bianchi identity that 
any acceptable curvature should satisfy.
And this constraint on the symmetry property of the additional ghosts 
is nothing but the nilpotence
property of the BRST symmetry.

In other words, the ordinary curvature is not a meaningful
geometrical object by itself in topological field theory,
rather it has to contain its own
ghosts representing the extra topological symmetry 
and this new curvuture has to satisfy the Bianchi identity 
in the extended space. 
Thus in the geometrical setting, the BRST symmetry of topological 
Yang-Mills-Higgs theory  is represented by the extended curvature
containing all ``permissible" ghosts, and the BRST symmetry of these
new ghosts are restricted by the Bianchi identity in the extended
space. I.e., from the BRST symmetry view point only the extended 
curvature containing all ``permissible" ghosts  is a geometrically 
meaningful object in topological Yang-Mills-Higgs theory. 
\\

\noindent
{\large \bf V. Conclusion } \\

In this paper, we found the rules for the BRST/anti-BRST symmetry
encompassing  topological Yang-Mills-Higgs theory in 
two, three and four dimemsions.
This was done in the superconnection framework so that 
the scalar field is regarded as a part of a connection 
as is the vector gauge field.
Using the superconnection language, we obtain the classical 
topological actions in two, three and four dimensions from a classical
action of topological Yang-Mills type through appropriate projections
depending on the dimensions of spacetimes to which 
corresponding theories belong. 
In this framework, the BRST/anti-BRST rules 
 for the scalar and vector gauge fields  are obtained together  
rather than separately. 
This also tells us that the usefullness of the superconnection language when
one deals with the scalar and vector gauge fields, since 
in the ordinary treatment the BRST/anti-BRST rules for these 
two fields are obtained separately. 
As a result, we extend the work of Perry and Teo \cite{pt} 
in the topological Yang-Mills case to the topological Yang-Mills-Higgs
case in two \cite{cg} and three \cite{bg} dimensions 
using the Bianchi identity. 
And comparing this work with our previous work of the modified 
horizontality condition approach, we conclude the following 
in topological Yang-Mills-Higgs theory: 
First, the newly defined ghost components of 
the extended curvature in the Perry and Teo's work 
can be identified as the objects representing the extra topological
symmetry which the theory possesses.
Second, the symmetry property that these new ghosts should obey is
constrained by the Bianchi identity in the extended space, and 
this requirement is nothing but the nilpotency condition of 
the BRST symmetry in another guise. 
Thus in theories with topological symmetry, 
it can be said that if things are expressed in the extended space which
contains the ghost directions, then one can treat the BRST symmetry 
in a geometrical setting in which the curvature contains 
all the ``permissible" ghosts, and the BRST symmetry 
due to topological symmetry is constrained by the Bianchi identity 
in this extended space.
\\ 

\noindent                                                             
{\large \it Acknowledgements} \\                                     
\indent                                                             
The author would like to thank D.S. Hwang for useful discussions.
This work was supported in part by NON DIRECTED RESEARCH FUND, 
Korea Research Foundation.


\begin{thebibliography}{99}                                           
          
\bibitem{wtt} E. Witten, Commun. Math. Phys. {\bf 117}, 353 (1988).
\bibitem{bs} L. Baulieu and I.M. Singer, Nucl. Phys. {\bf B 5B}, 12 (1988).
\bibitem{pt} M.J. Perry and E. Teo, Nucl. Phys. {\bf B 392}, 369 (1993).
\bibitem{brst} C. Becchi, A. Rouet, and R. Stora, Phys. Lett. {\bf B 52},
           344 (1974); Comm. Math. Phys. {\bf 42}, 127 (1975); Ann. Phys.
          {\bf 98}, 287 (1976); 
           I.V. Tyutin, Lebedev report FIAN No. 39 (1975).
\bibitem{cf} G. Curci and R. Ferrari, Nuovo Cimento {\bf 30 A}, 155 (1975);
              T. Kugo and I. Ojima, Suppl. Prog. Theor. Phys. {\bf 66}, 1
                (1979).
\bibitem{bg} L. Baulieu and B. Grossman, Phys. Lett. {\bf B 214}, 223 (1988).
\bibitem{cg} G. Chapline and B. Grossman, Phys. Lett. {\bf B 223}, 336 (1989).
\bibitem{hl2} D.S. Hwang and C.Y. Lee, J. Math. Phys. {\bf 37}, 693 (1996).
\bibitem{tmn} J. Thierry-Mieg and Y. Ne'eman, Proc. Natl. Acad. Sci. USA
  {\bf 79}, 7068 (1982).
\bibitem{crt} E. Cartan, C. R. Acad. Sci. Paris {\bf 182}, 956 (1926).
\bibitem{qui} D. Quillen, Topology {\bf 24}, 89 (1985).
\bibitem{ns} Y. Ne'eman and S. Sternberg, Proc. Natl. Acad. Sci.
              USA {\bf 87}, 7875 (1990).
\bibitem{bz} F.B. Berezin, {\it Introduction to Superanalysis}, 
           (Reidel Pub., Dordrecht, 1987), Chap. 3.
\bibitem{tm} J. Thierry-Mieg, J. Math. Phys. {\bf 21}, 2834 (1980);
          Nuovo Cimento {\bf 56 A}, 396 (1980).
\bibitem{ntm} Y. Ne'eman and J. Thierry-Mieg,
  in {\it Proceedings of the 1979
  Salamanca International Conference on Differential Geometric
  Methods in Mathematical Physics (1979)}, edited by  P.L.
  Garcia, A. Perez-Rendon and J.M. Souriau,
  Lecture Notes in Mathematics No. 836 (Springer-Verlag, Berlin,
  1980);  Proc. Natl. Acad. Sci. USA
  {\bf 77}, 720 (1980).
\bibitem{qrs}  M. Quiros, F. J. De Urries, J. Hoyos, M. L. Mazou, and
             E. Rodriguez,  J. Math. Phys. {\bf 22}, 1767 (1981).
\bibitem{tmn2}  J. Thierry-Mieg and Y. Ne'eman, Nuovo Cimento 
     {\bf 71 A}, 104 (1982).
\bibitem{btm} L. Baulieu and J. Thierry-Mieg, Nucl. Phys. {\bf B 197},
      477 (1982); {\sl ibid.} {\bf B 228}, 259 (1983).
\bibitem{ln} C.Y. Lee and Y. Ne'eman, Phys. Lett. {\bf B 264}, 389 (1991);
      Erratum  {\bf B 269}, 477 (1991).
\bibitem{botn} L. Bonora and M. Tonin, Phys. Lett. {\bf B 98}, 48 (1981).
\bibitem{bl}  L. Baulieu, Phys. Rep. {\bf 129}, 1 (1985).
\end{thebibliography}
\end{document}